\documentclass[journal]{IEEEtran}
\usepackage{cite}
\usepackage{xcolor}
\usepackage{amssymb}
\usepackage{graphicx}
\usepackage[utf8]{inputenc}
\usepackage{dirtytalk}
\usepackage{soul}
\ifCLASSINFOpdf

\else
 
\fi

\begin{document}

\title{Experimental Side Channel Analysis of BB84 QKD Source}

\author{Ayan~Biswas,%~\IEEEmembership{Member,~IEEE,}
        ~Anindya~Banerji,%~\IEEEmembership{Fellow,~OSA,}
        ~Pooja~Chandravanshi,%~\IEEEmembership{Life~Fellow,~IEEE}%
         ~Rupesh~Kumar,~%~\IEEEmembership{Life~Fellow,~IEEE}
          and Ravindra~P.~Singh%~\IEEEmembership{Life~Fellow,~IEEE}%% stops a space
\thanks{A. Biswas, P. Chandravanshi and R.P. Singh are with the Quantum Technologies Laboratory, Physical Research Laboratory, Ahmedabad, 380009 India. (e-mail: rpsingh@prl.res.in)}% <-this % stops a space
\thanks{A. Banerji is with Centre for Quantum Technologies, National University of Singapore, 3 Science Drive 2, Singapore 117543. (e-mail: cqtab@nus.edu.sg)}% <-this % stops a space
\thanks{A. Biswas is also with Indian Institute of Technology, Gandhinagar, 382355 India. (corresponding author: ayan@prl.res.in)}% <-this % stops a space
\thanks{R.K. is with Quantum Communications Hub and York Centre for Quantum Technologies, Department of Physics, University of York, York, YO10 5DD, UK. (e-mail: rupesh.kumar@york.ac.uk)}% <-this % stops a space

}


\maketitle

\begin{abstract}
A typical implementation of BB84 protocol for quantum communication uses four laser diodes for transmitting weak coherent pulses, which may not have the same characteristics. We have characterized these lasers for mismatch in various parameters such as spectral width, pulse width, spatial mode, peak wavelength, polarization and their arrival times at the receiver. This information is utilized to calculate possible information leakage through side channel attacks by evaluating mutual information between source and eavesdropper. Based on our experimental observations of cross correlation between
parameter values for different laser diodes, we suggest methods to reduce  information
leakage to Eve.  
\end{abstract}

\begin{IEEEkeywords}
Quantum key distribution, Side channel, Cross-correlation, Quantum communication,
\end{IEEEkeywords}

\IEEEpeerreviewmaketitle

\section{Introduction}

\IEEEPARstart{Q}{uantum} Cryptography \cite{10.1145/1008908.1008920,BENNETT20147} is gaining importance with the advancements in quantum computation which can make the present cryptographic techniques redundant. It exploits the basic principles of quantum mechanics to generate or distribute a secret key between two communicating parties. This process is referred as quantum key distribution (QKD) \cite{BENNETT20147}. The goal is to ensure that message remains confidential and inaccessible to a third party. 
% With more involvement of online transactions and activities it becomes important to keep the secrecy between the sender and the receiver from foreign elements. 
\par The first QKD protocol was proposed by Bennett and Brassard in 1984 \cite{BENNETT20147} and referred as the BB84 protocol. In this protocol, the secret key bits are encoded in the quantum states, for example, polarization of a single photon. \textcolor{black}{The states to be sent are prepared in mutually unbiased bases (MUB)}.The intrinsic uncertainty in measurement of polarization in randomly selected MUB \cite{10.1145/1008908.1008920,BENNETT20147,Bennett1992} makes this protocol secure in principle. The security proof of the BB84 protocol came much later \cite{shor2000simple,Lo2050}, which was followed by a security analysis for implementation scenarios with imperfect devices \cite{gisin2002quantum,Gottesman:2004:SQK:2011586.2011587,Mayers2002}. \textcolor{black}{There are QKD protocols, other than BB84, which use degree of freedom different from polarization. For example COW (Coherent One Way), SARG04 use phase instead of polarization to encode the states \cite{PhysRevLett.92.057901,Stucki_2009}. These protocols lag behind on efficiency and easiness relative to BB84 in terms of practical implementation. All these protocols use various devices which are not perfect in reality and can be prone to attacks \cite{jain2014risk,jain2014trojan,lydersen2010hacking,huang2018implementation,shenoy2017quantum}. To reduce the vulnerabilities in the QKD protocols due to imperfect measurement devices, MDI (Measurement Device Independent) QKD schemes have been devised \cite{PhysRevLett.108.130503,Lucamarini2018,PhysRevX.6.011024,PhysRevLett.117.190501,Sci.Rep.9}, however, their implementations are much more difficult in practice. On the other hand BB84 has solid theoretical security backup \cite{renner2008security,RevModPhys.81.1301} against a wide range of attacks and widespread demonstrations around the world, which makes it a preferred protocol to apply in practice}. 
%There are many security proofs for this kind of QKD protocol but, most of them are based on theoretical considerations \cite{renner2008security,RevModPhys.81.1301}. However in, implementation this may vary from scheme to scheme and also with the devices that one is using. 
%As the devices itself have imperfections \cite{jain2014risk,jain2014trojan,lydersen2010hacking,huang2018implementation,shenoy2017quantum}which might open doors for guessing the key by an adversary. 
\par Even though there are several security proofs based on the attack strategies of Eve but most of them assume the devices and the optical elements used in the QKD setup to be ideal. There is a possibility that Eve might know the weakness of these devices and can work this out to her benefit in guessing the secret key. This so called side channel attack \cite{ko2017critical,pirandola2019advances} is very serious in any QKD system as Eve can get information about the key directly from the devices that are being used. One way to get rid of this is to use fully device independent QKD, however, it is still not widely used  \cite{PhysRevLett.113.140501,PhysRevLett.108.130503,Lucamarini2018} since they are much more resource intensive unlike general prepare and measure protocol such as BB84, at the same time their key generation rate is quite low. Although one cannot avert side channel attack completely but can make sure that Eve's gain is minimized. This information leakage \cite{nauerth2009information,jain2016attacks,sajeed2019approach} to Eve will keep an upper bound on the key after error correction and privacy amplification (PA) \cite{10.1137/0217014}. For properly quantifying the amount of information going to Eve, one needs to know the limitations in the devices that are used in the QKD process. 
%Here, we have only calculated the side channel information to Eve only at the source (Alice's) part. 
\textcolor{black}{Devices at both ends (Alice and Bob) must be calibrated in order to quantify the side information to Eve. Device imperfections at the Bob's end are easy to measure as these can be verified through parameter estimation for QKD}. 
\par \textcolor{black}{In the present article, we are interested in calculating Alice's information leakage as the vulnerabilities associated with the source are highly prone to leakage. This is because signals passing through the quantum channel can always be under Eve’s surveillance. Therefore, the primary concern is to evaluate the various source parameters for side channel attack. For quantifying this, one has to calculate the mutual information between the source and Eve. The inability of the source to produce ideal states for QKD will give information to Eve. This can depend upon various parameters characteristics to the source. To know the amount of information leakage, one has to meticulously calibrate the source for these parameters, which could be wavelength, photon arrival time, or any other parameter making the transmitted states distinguishable from each other such that Eve can easily gain some information out of it. Therefore, source calibration is essential for knowing the mutual information between the source and the adversary. \textcolor{black}{In this article, we calibrate our QKD transmitter consisting of four laser diodes through cross correlation for various parameters and use it to estimate the mutual information \cite{shannon1948mathematical,gisin2002quantum} between Eve (E) and Alice (A), unlike \cite{nauerth2009information} that calculates conditional probabilities to find out this information}. The work is divided into four sections. The first section describes the basics of the method for calculating the mutual information in our experiment. The experimental method used for this is elaborated in the second section. In the third section, we discuss the results of the experiment performed. In the last section, we conclude our work with suggestions to minimize the leakage to Eve through the side channel.}

 

\section{Theoretical Background}
Mutual information is basically the correlation between variables of the two parties involved in communication. It quantifies the amount of knowledge one has about the other. The mutual information between two parties Alice (A) and Bob (B) 
%with associated random variables $X= \{ x_{1},x_{2},x_{3},... \}$ and $Y=\{ y_{1},y_{2},y_{3},... \}$ 
can be written as 
\cite{shannon1948mathematical,gisin2002quantum,renner2008security}
\begin{equation}
    I(A:B)=H(A)-H(A|B).
\end{equation}
$H(A)$ is the information entropy Alice has about her variables and $H(A|B)$ is the conditional entropy of Alice given Bob measured his. This can be further simplified in terms of corresponding probabilities  
\begin{equation}
    I(A:B)=H(A)+\sum_{a \epsilon A}p(a)\sum_{b \epsilon B}p(b|a)Log(p(b|a)),
\end{equation}
where $p(a)$ is the probability of occurrence of an event \say{$a$} and $p(a|b)$ is probability of happening \say{$a$} given the event \say{$b$} has already occurred. The event \say{$a$} can be an outcome from the space $A$, similarly \say{$b$} can be an outcome from the space $B$ and \textcolor{black}{$Log$ is taken in  base 2}. 
In practice this tells us how much Bob's \textcolor{black}{information} is correlated to Alice. The secret key rate for reverse and direct reconciliation in QKD protocol is given by \cite{gisin2002quantum,pirandola2019advances,RevModPhys.81.1301}
\begin{equation}
    r_{DR} \geqslant I(A:B)-I(A:E)
\end{equation}
\begin{equation}
    r_{RR} \geqslant I(A:B)-I(B:E)
\end{equation}
where subscript $RR$ means reverse reconciliation \textcolor{black}{in which Alice corrects the erroneous bits after sifting by comparing it with the Bob's key}. $DR$ means direct reconciliation \textcolor{black}{where Bob makes correct changes in the key by verifying it with Alice after sifting}. For secure QKD protocol this quantity should be non zero. For $r_{DR},r_{RR} \geqslant 0$ implies that the mutual information between Alice and Bob ($I(A:B)$) must be greater than Alice and Eve ($I(A:E)$) and Bob and Eve ($I(B:E)$). $I(A:E)$ can be found out if we know the transmission channel and also the errors occurring at the source and $I(B:E)$ can be measured by looking at the imperfections in the detection unit setup by Bob. If the two parties happen to be Alice and Eve then Eq (2) can be rewritten as \cite{nauerth2009information,pirandola2019advances} 
\begin{equation}
     I(A:E)=1+\sum_{a \epsilon A}p(a)\sum_{b \epsilon E}p(b|a)Log(p(b|a))
\end{equation}
\textcolor{black}{In equation (5), $b$ is the outcome of Eve's measurement in her device. Given the outcome $a$ at Alice’s side, Eve's probability to measure bit $b$ is the conditional probability ($p(b|a)$). In reality, Alice doesn’t know what method Eve will use; therefore, for quantifying this information, the measurement has to be done from Alice’s side. It means she has to make sure how much Eve can guess about the states she is sending. To calculate the conditional probability for a variable, we use Bayes’ Theorem by taking the source’s parameter that deviates from the assumed value in the implemented protocol. Since we have used four different laser diode sources, due to differences in electronic fluctuations in the driving circuit, there may be a difference in wavelength ($\lambda$), pulse width ($w$), and other parameters. Equation (5) decides the amount of information leakage to Eve due to imperfections in these parameters of the source. To characterize this quantity, we need to measure the amount of indistinguishability between various source parameters.} 
\par \textcolor{black}{In Eq (5), the primary quantity that needs to be calculated is the conditional probability $p(b|a)$, as this quantity decides the amount of information shared between Alice and Eve. Here, the parameters in consideration are wavelength ($\lambda$), pulse width ($w$), photon time arrival ($t$), polarization error at the source, and spatial mode ($x$). These are the events occurring in Alice’s system and the bit value that Eve gets after measuring the states is $b$ (it can either be 0 or 1). $p(\lambda|b)$ can be calculated with the help of joint probability distribution ($p(\lambda , b)$). We can rewrite Eq (5) in terms of experimental parameters}  
\begin{equation}
     I(A:E)=1+\sum_{\lambda \epsilon \Lambda}\sum_{b \epsilon E}\frac{p(\lambda|b)}{2}Log\Big(\frac{p(\lambda|b)}{2p(\lambda)}\Big)
\end{equation}
Here, $\lambda$ is the wavelength of the laser having a finite bandwidth (FWHM). $\Lambda$ is the space containing all values of $\lambda$. \textcolor{black}{Just as wavelength ($\lambda$), Eq (6) will be identical for other parameters also}. For pulse width $ \Lambda $ is replaced by $W$, similarly for photon time arrival it is $T$ and $X$ for spatial mode. Estimating $I(A:E)$ as given in Eq (6) can be slightly time taking task if we are implementing the QKD source in the field. Instead we have come up with a method that can quickly give us the amount of information leaked to Eve quantitatively. We, need to calculate the cross correlation ($R$) between various parameters of the two sources of the same basis in the QKD transmitter to quantify the amount of information leakage. The expression for cross correlation is given by \cite{10.5555/1841670}
\begin{equation}
    R(\Delta s)= \int f^{*}(s)g(s+\Delta s) ds.
\end{equation}
Equation (7) tells us about the similarity between the two signals $f$ and $g$ as a function of $\Delta s$. The quantity $\Delta s$ is the shift of one signal with respect to other and the value of $R$ ranges from 0 to 1. The indistinguishability between the sources can be known from $R(\Delta s=0)$ (more close to 1 means more similar to each other). Putting $R(\Delta s=0)$ in the Eq (6) we get
\begin{equation}
    I(A:E)=1+\sum\frac{R(0)}{4}Log\Big(\frac{R(0)}{4p(\lambda)}\Big).
\end{equation}
$R$ is the measure of offset of the parameters like wavelength, pulse width etc. between the four laser diodes. It is impossible for Eve to predict the state of Alice's signal after measurement if the parameters of the sources are identical. $p(\lambda|b)$ tells about probability of guessing the correct initial state after the measurement by Eve. This value is half 
%for exactly two identical signals from the laser diodes 
i.e upon getting a bit value 1 it is impossible to say whether it comes from source containing $\lambda_1$ or $\lambda_2$. Deviation from this value basically gives us the quantity of leaked information. $R(\Delta s =0)$ represents the deviation from indistinguishability between the various parameters of the source. By argument, one can say that the quantity $R (\Delta s =0) \times \frac{1}{2}$ is the guessing probability of Eve, so this quantity can be replaced with $p(\lambda|b)$. For exactly identical states, the quantity $I(A:E)$ will be zero as the parameters of the source are indistinguishable which can be verified using Eq (8). \textcolor{black}{Therefore, measuring the cross correlation between the various parameters gives a good idea about amount of information leakage to the eavesdropper.}
\section{Experimental Method}
The schematics of the experiment is given in the figure 1
\begin{figure}[h!]
\includegraphics[height=3.4cm,width=8.8cm]{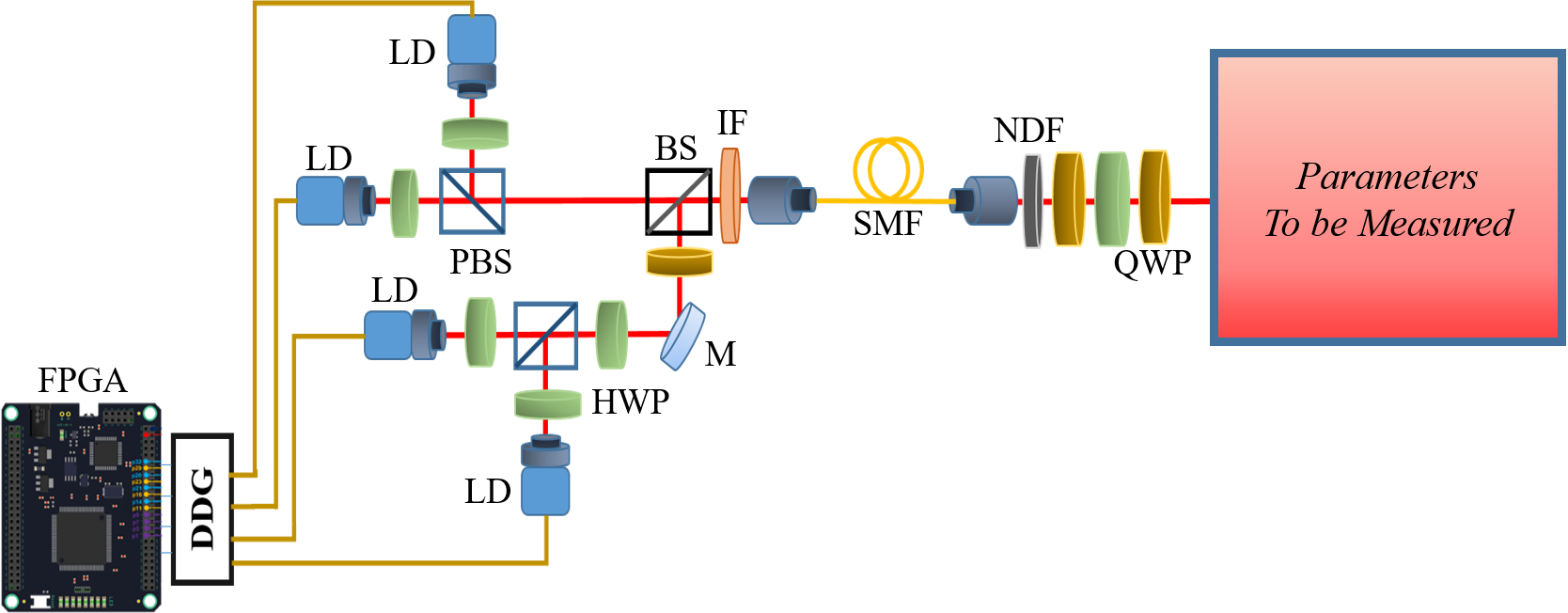}
\caption{Experimental scheme for measuring the parameters involved in source characterization. For different parameters one has to change the measuring devices. \textbf{LD}: Laser Diode, \textbf{FPGA}: Field Programmable Gate Array, \textbf{PBS}: Polarizing Beam Splitter, \textbf{HWP}: Half Wave Plate, \textbf{QWP}: Quater Wave Plate, \textbf{M}: Mirror, \textbf{SMF}: Single Mode Fiber, \textbf{NDF}: Neutral Density Filter, \textbf{IF}: Interference Filter, \textbf{DDG}: Digital Delay Generator, \textbf{BS}: Beam Splitter.}
\end{figure}

In the setup, the measurement devices can be changed according to the parameters that need to be measured for the experiment. The scheme contains four laser diodes (ThorLabs L808P010) with driver circuit \cite{10.1117/12.545038}. %\textcolor{red}{Like for wavelength optical spectrometer is used, for pulse width measurement that is replaced by photodetector and oscilloscope, for photon time arrival single photon detector with time tagger is used at the output and for spatial mode EMCCD is used}. 
The pulse width coming out from the laser can be varied according to input bias voltage. We keep the average pulse width around (650 ps) and the repetition rate of the laser is 5 MHz. The laser driver circuit is connected to stable power supply (Keithley 2231A-30-3) and FPGA (Arty A7) for driving it randomly. The initial HWP and PBS combination is used for preparing specific polarization states for encoding Alice's signal. All of the four sates coming out from the laser diodes are combined in the 50:50 beam splitter and coupled to a single mode fiber (SMF). \textcolor{black}{The SMF is used to reduce any sort of misalignment error in the four lasers}. Then a combination of QWP, HWP and QWP is used for compensating the polarization after propagation through the fiber \cite{Reddy:14}. For measuring wavelength, we just place an optical spectrometer (Ocean Optics HR4000), for pulse width it is a fast photodetector whose output is connected with oscilloscope for monitoring the signal. For measuring the spatial mode we use EMCCD camera. Photon Arrival time is measured by placing a single photon counting module (Excelitas SPCM-AQRH-16) connected to time tagger (ID Quantique ID900). The clock of frequency equal to the driving frequency of the laser is sent to TDC for starting the counting time of photon arrival. The histogram will give us the knowledge of time of arrival of photons with respect to clock.
\section{Results and Discussions}
 Experimental results show that the proper source characterization should be done to quantify the amount of information leakage due to side channel attack by the adversary. The following parameters have been quantified for indistinguishability between the individual laser diodes of the BB84 source.  

\subsection{Wavelength}
The mismatch between the peak wavelengths for the laser diodes can give eavesdropper a chance of differentiating between different polarization states by looking at the mismatch between the wavelengths. The figure shows the wavelengths \textcolor{black}{(nm) verses normalised intensity in terms of counts per second (cps)} for four laser diodes and their mismatch in terms of peak difference.

\begin{figure}[h!]
\includegraphics[height=6.5cm,width=8.6cm]{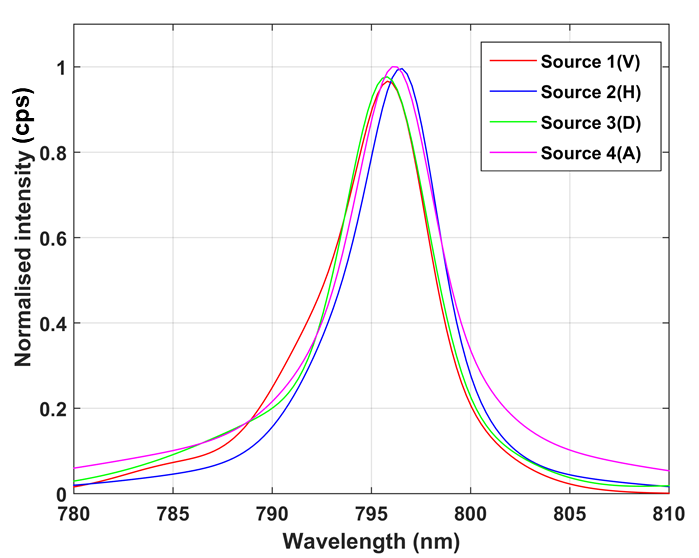}
\caption{Spectrum of four laser diodes without using Interference Filter (IF)}
\end{figure}
The source is having the average wavelength of 795.6 nm.%, with standard deviation of $\pm$ 0.28 nm. %The resolution of the optical spectrum analyzer is 0.25nm. 
The measurements have been taken without putting any wavelength filters. Figure shows the wavelength of the four laser diodes in the Alice transmitter unit.
%\begin{figure}[h!]
%\includegraphics[height=3cm,width=8.5cm]{WMM.png}
%\caption{Setup for measuring the wavelengths from four laser diodes}
%\end{figure}
The information leakage due to wavelength difference between the four laser diodes is calculated from the cross correlation between the lasers is $I_{\{H,V\}}(\Lambda : E)=4.3 \times 10^{-3}$ \textcolor{black}{bits/pulse} where the subscript $H,V$ denotes the information leakage in H/V basis and $\Lambda$ and $E$ are the corresponding spaces on which a typical $\lambda$ and $b$ belong, similarly $I_{\{D,A\}}(\Lambda : E)=6.5 \times 10^{-3}$ \textcolor{black}{bits/pulse} giving \textcolor{black}{mutual information} $I(\Lambda : E) \propto 10^{-3}$ \textcolor{black}{bits/pulse}.      

\subsection{Pulse width} 
The difference between the FWHM of the pulses (pulse width) from the laser diodes can give eavesdropper a chance of differentiating the transmitted states. In our setup, the shape of the RNG output pulses from the FPGA that are fed into the driver circuit are identical but, the optical response of the four laser diodes is not completely identical leading to a difference in the pulse widths from each diode. Therefore, optical output pulse of the laser diode is independent of the RNG pulses fed into FPGA. This variation in the pulse width creates some degree of distinguishability which can be exploited by Eve. Eve can unambiguously detect the polarization states sent from Alice just by looking at their pulse width variation in her detector. For characterizing this error the measurement scheme is modified by replacing spectrometer with photodetector.
%\begin{figure}[h!]
%\includegraphics[height=3cm,width=8.5cm]{PWMM.png}
%\caption{Experimental setup for measuring the pulse width of the four laser diodes}
%\end{figure}
The source has an average pulse width of 627 $\pm$ 75 ps as shown in figure 3 . \textcolor{black}{In QKD mode (sending qubits to Bob) pulse height is made identical by applying different attenuation to different states. All the four sources then have identical height hence having same mean photon number ($\mu$)}.
\begin{figure}[h!]
\includegraphics[height=6.9cm,width=8.5cm]{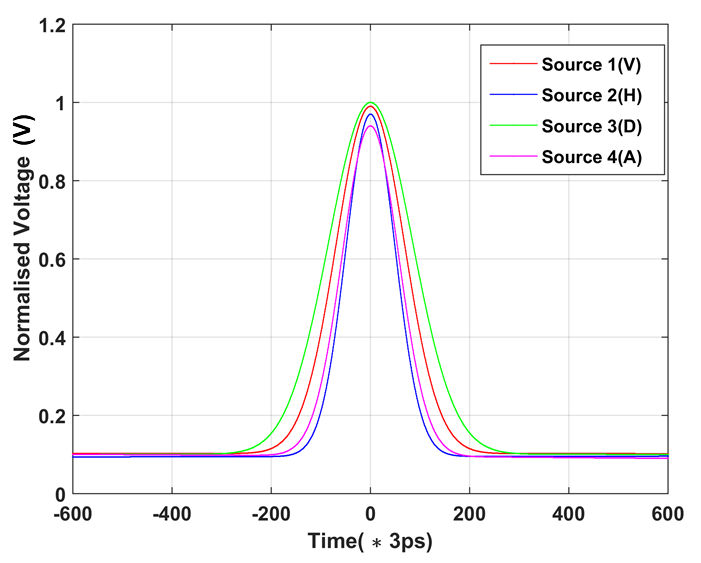}
\caption{Pulse width of four laser beam coming out from different laser drivers}
\end{figure}
The information leakage due to pulse width in laser diodes calculated from the cross correlation between the sources is $I_{\{H,V\}}(W:E)=9.2 \times 10^{-4}$ \textcolor{black}{bits/pulse}, $I_{\{D,A\}}(W:E)=1.2 \times 10^{-3}$ \textcolor{black}{bits/pulse} with \textcolor{black}{mutual information} $I(W:E)\propto 10^{-3}$ \textcolor{black}{bits/pulse}.
\par Eve can design optimized attacking strategies based on the knowledge of both wavelength and pulse width from which she
can learn more about the state. In fact, Eve can also exploit all the side channels together which may allow her to extract more information. However, it is too complex to conceive a best attack strategy which is out of scope of this article. Nevertheless, we will try to consider it in our future work.

\subsection{Arrival time}
The arrival time of photons will depend on at what time the photons from different laser diodes are leaving the Alice's QKD transmitter. The difference in the initial timing will give a hint to eavesdropper about the corresponding states being sent to Bob. Even if the optical circuit is perfect, the electrical driver circuit which triggers the on and off time of the laser diodes is subject to jitter. This will result in pulses from different diodes leaving the transmitter at different times causing a difference in the photon arrival times. This can be exploited by Eve to extract information about the states sent from Alice to Bob by looking into the timing information that is disclosed during the sifting stage in the QKD protocol \cite{Lamas-Linares:07}. In order to know the amount of information that can be gained by Eve, one has to measure photon arrival time. For measuring it, the attenuated pulses need to be sent to single photon detector and the output of that is taken from a time counter.
%\begin{figure}[h!]
%\includegraphics[height=3cm,width=8.5cm]{TMM.png}
%\caption{Experimental scheme for photon arrival time measurement}
%\end{figure}
The average time of arrival of the photons is almost same for the four laser diodes as seen in the figure 4 which in this case is 41.34 $\pm$ 0.075 ns. The information leakage due to photon arrival time difference for four laser diodes calculated from the cross correlation between the sources is $I_{\{H,V\}}(T:E)=3.2 \times 10^{-3}$ \textcolor{black}{bits/pulse}, $I_{\{D,A\}}(T:E)= 2.5 \times 10^{-3}$ \textcolor{black}{bits/pulse} with \textcolor{black}{mutual information} $I(W:E)\propto 10^{-3}$ \textcolor{black}{bits/pulse}.

\begin{figure}[h!]
\includegraphics[height=6.9cm,width=8.5cm]{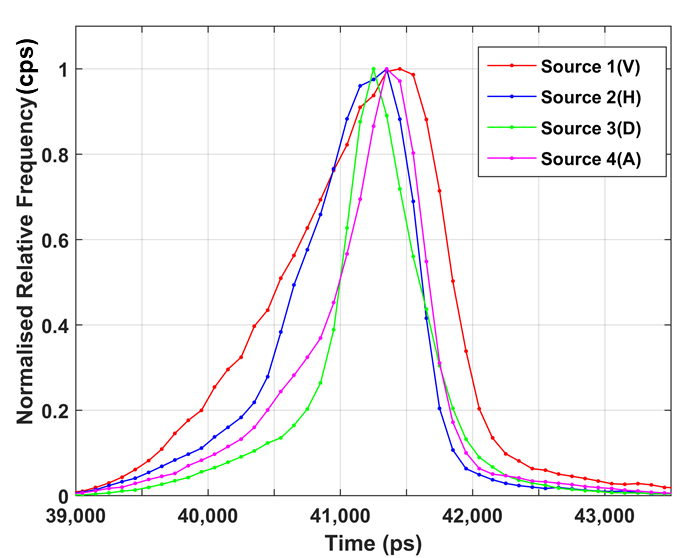}
\caption{Graph showing arrival time of photons from four different laser diodes}
\end{figure}

\subsection{Polarization Error at Source}
The use of optical devices in the transmitter to combine the beams from four diodes into one may lead to many imperfections. These imperfections may occur either due to misalignment in the transmitter setup or due to imperfect optics. Therefore, the generated states are not perfect in terms of polarization and may contain error which may reflect in the polarization extinction ratio in H,V as well as D,A basis. This error may lead to information leakage to Eve. It gets further amplified in the final QBER after the states are sent to Bob. Figure 5 gives the errors in polarization.
\begin{figure}[h!]
\includegraphics[height=7cm,width=9.5cm]{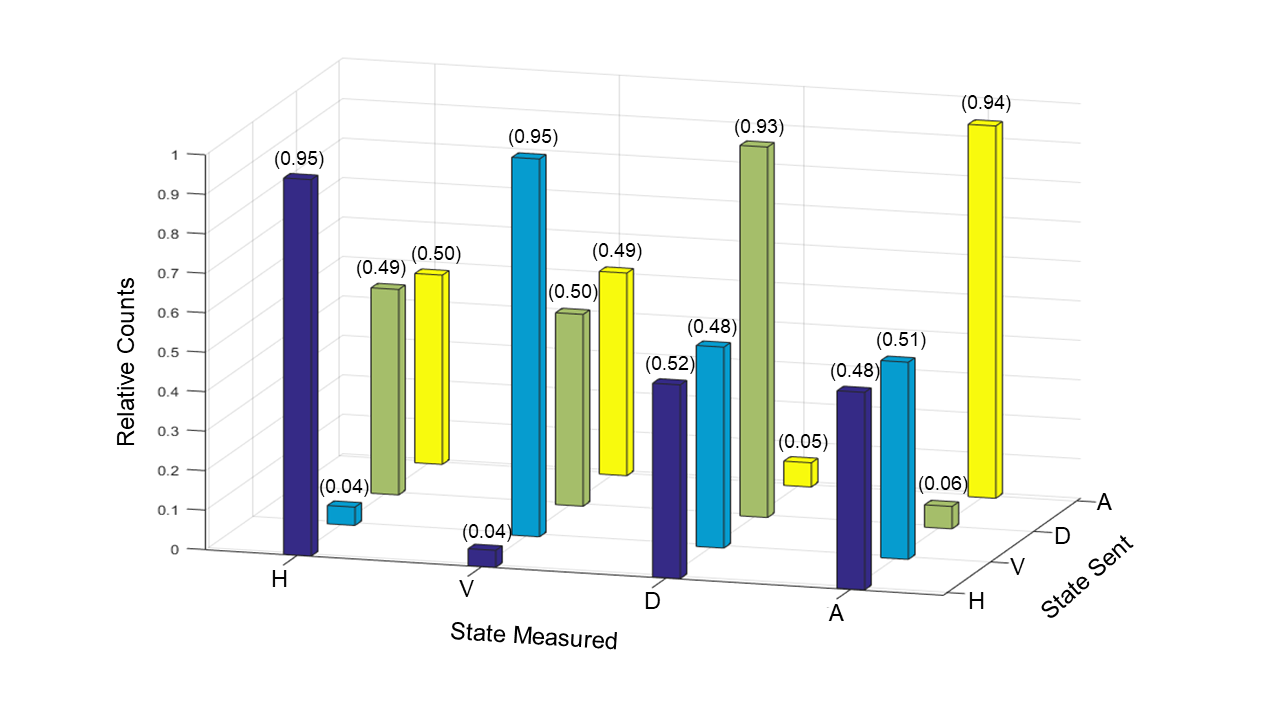}
\caption{Polarization error at the source}
\end{figure}
%\begin{table}[h!]
%\centering
%\begin{tabularx}{0.4\textwidth} { | %>{\raggedright\arraybackslash}X | %>{\centering\arraybackslash}X | 
%>{\centering\arraybackslash}X |
%>{\centering\arraybackslash}X |
%>{\centering\arraybackslash}X |}
%   \hline
 %  & H & V & D & A \\
 %  \hline
 %  H & 95.35 $\%$ & 4.65$\%$ & 52.05$\%$ & 47.95 $\%$ \\
 %  \hline
 %  V  & 4.30$\%$ & 95.7$\%$ & 48.6$\%$ & 51.34$\%$  \\
%   \hline
 %   D  & 49.09$\%$ & 50.91$\%$ & 93.86$\%$ & 6.14$\%$  \\
 %  \hline
 %   A  & 50.05$\%$ & 49.95$\%$ & 5.67$\%$ & 94.33$\%$  \\
 %  \hline
%\end{tabularx}
%\caption{Polarization error at the source}
%\label{table:1}
%\end{table}
It shows that from mismatch in the basis dependent error (in non-compatible basis) Eve can extract information about the bits sent by Alice. Error in $H/V$ and $D/A$ basis are $e_{H/V}=0.0341$ and $e_{D/A}=0.0094$ respectively and their mismatch is $\Delta e = |e_{H/V}-e_{D/A}|$. While doing basis reconciliation in QKD Eve can guess the bits sent to Bob by the data she already had about this mismatch. So, information shared between Alice and Eve is $I(A:E) \propto \Delta e \propto 10^{-2}$ \textcolor{black}{bits/pulse}.     

\subsection{Spatial Mode}
While doing free space QKD it becomes very important to look at the modes of the signal that are propagating through the medium. The spatial mode may be responsible for creating vulnerability in the QKD source. If the modes do not perfectly overlap with each other it may hamper the indistinguishability of four quantum sates. \textcolor{black}{If the four beams enter the fiber with different injection angles then the output spot size distribution will be different for different beams and is evident while using shorter length fibre for mode cleaning}. 
%The spot sizes may vary more for multi-mode fiber (MMF) as the core diameter is 30-150 $\mu m$, compared to SMF with core diameter of 5-15 $\mu m$}. 
This mismatch in spatial modes can be measured by EMCCD camera. For making four spatial modes overlap with each other one needs to couple them into a short length single mode fiber. Earlier work has not given much emphasis on mismatch among spatial modes as they amounted to very less leakage \cite{nauerth2009information} but here we show that if Eve has a very low pixel size camera then she could measure the mode mismatch in four laser diodes. Experimental scheme remains the same as in figure 1 only this time EMCCD camera is used instead of spectrometer.
%\begin{figure}[h!]
%\includegraphics[height=3cm,width=8.5cm]{SMMM.png}
%\caption{Setup for recording spatial mode mismatch between four laser diodes}
%\end{figure}
 %After taking the images of the spatial profile of the four laser diodes one by one and recording in EMCCD camera having pixel size of $16\mu m \times 16 \mu m$. 
 Images of spatial modes of four laser diodes are recorded in EMCCD camera. Figure 6 shows the spatial mode distribution of these four laser diodes.
\begin{figure}[h!]
\includegraphics[height=5.5cm,width=8.9cm]{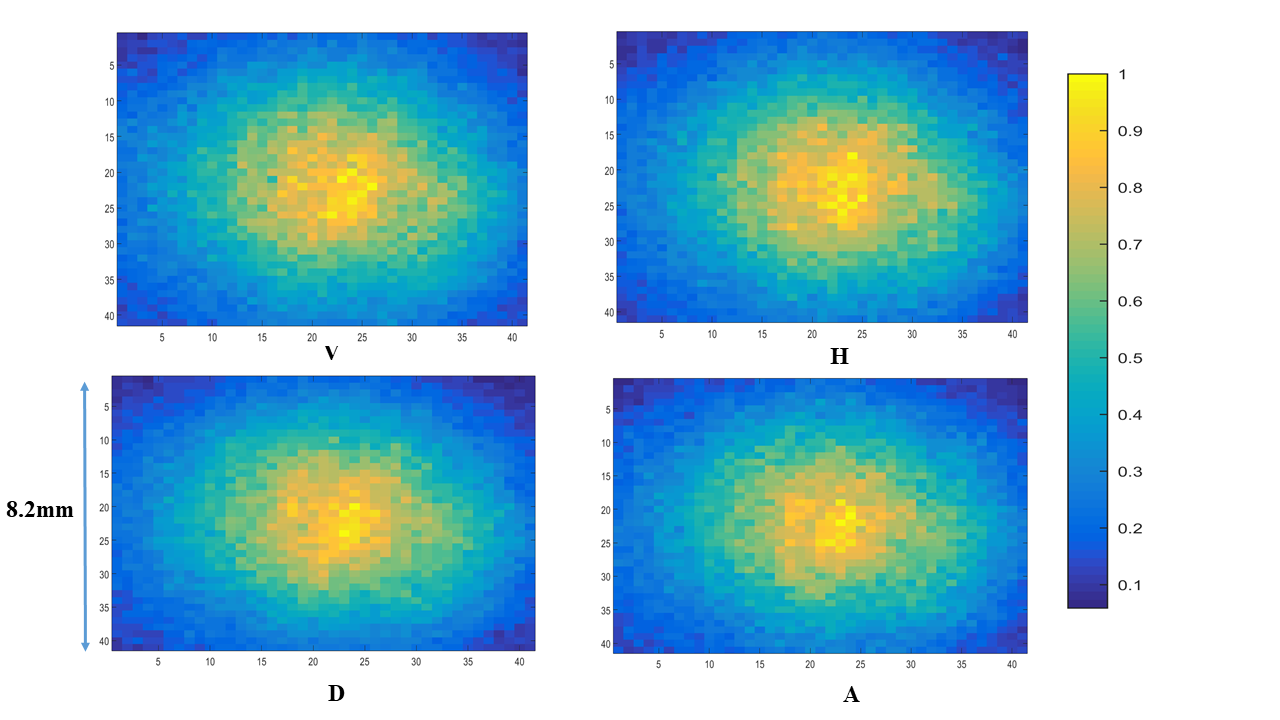}
\caption{Images of beams taken at the out of the fiber which are coming from four laser diodes}
\end{figure}
The information leakage due to spatial mode difference for the four laser diodes calculated from the cross correlation between them is $I_{\{H,V\}}(X:E)=4.2 \times 10^{-3}$ \textcolor{black}{bits/pulse}, $I_{\{D,A\}}(X:E)=4.5 \times 10^{-3}$ \textcolor{black}{bits/pulse} with \textcolor{black}{mutual information} $I(X:E)\propto 10^{-3}$ \textcolor{black}{bits/pulse}.
\par \textcolor{black}{The highlight of the present work which makes it different from \cite{nauerth2009information} is the consideration of new parameters that can also contribute to the information leakage. Pulse width variation is an important parameter in the source which gives rise to side channel information to the adversary. Our work quantifies this leakage of information to Eve due to pulse width mismatch between four laser diodes. Secondly, the polarization error in different bases at the source is also an important quantity which sets a bound in the side channel information to Eve. Lastly, it has been shown that leakage can take place if Eve uses detector with smaller pixel size to find the spatial mode distribution of four states. The smaller the pixel size, the better the Eve will be able to discriminate four spatial modes with higher certainty. However, \cite{nauerth2009information} concludes that spatial measurements leads to negligible information leakage.} 
\par\textcolor{black}{In the given reference \cite{nauerth2009information}, to find out the information leakage, laborious way of calculating the conditional probabilities has been used. Instead we use the cross correlation technique, which is much more experiment friendly, simple and also gives a good estimation of the side channel leakage in the QKD source.}
\section{Conclusion}
\textcolor{black}{In the present work we have characterized the various source parameters that can lead to possible side channel attack. Using cross correlation function for calculating mutual information between Alice and Eve gives quite good results. This method is simple and can be easily implemented in the field which can give real-time values for possible information leakage. In our setup parameters such as pulse width and spatial mode contributes less to side channel information to Eve}. The information leakage once quantified can be crucial for extracting the secret key after privacy amplification. This mutual information can further be decreased by correcting the optical and the electronic elements in the source. To make a better BB84 source, one has to make sure that the states that are being created must be as indistinguishable as possible. This makes very difficult for Eve to guess the states correctly by knowing the parameters of the source. For wavelength mismatch one can put very narrow bandwidth filters which can decrease the the overall FWHM as well as the peak to peak mismatch between the four laser diodes. \textcolor{black}{Using precise temperature control and stabilization methods one} can reduce the wavelength mismatch between the four laser diodes. For making the pulse width of the laser diodes same, one can build a common laser driver circuit for them. For making arrival times same for all the four laser diodes one has to put fast delay generator in the driving circuit. For removing spatial mode mismatch, one can use long length fiber for mode cleaning. For reducing polarization error, one can use polarization maintaining fiber and broad-band optical elements in the setup. The parameters mainly contributing to information leakage are wavelength and polarization errors which need special attention while developing the QKD source. Earlier spatial mode was thought to be contributing less in information leakage \cite{nauerth2009information,nauerth2013air} but if Eve has good resolution camera, then she can guess the states with more confidence. Pulse width mismatch contributes less in this leakage and can be overlooked if the bit length of the secure key is not the matter of concern.

%\appendices

%\section{}

\section*{Acknowledgment}
\textcolor{black}{This work is partially funded by DST through QuST program. R. K. acknowledges the support from UK EPSRC through Quantum Technology Hub for Quantum Communications Technology, grant no. EP/T001011/1.}

\ifCLASSOPTIONcaptionsoff
  \newpage
\fi

%\end{IEEEbiography}



\end{document}